\UseRawInputEncoding
\documentclass[%
 reprint,
superscriptaddress,
 amsmath,amssymb,
 aps,
prb,
]{revtex4-1}

\usepackage{graphicx}
\usepackage{dcolumn}
\usepackage{bm}
\usepackage{amsmath,amssymb}
\usepackage{graphicx}
\usepackage{color}

\usepackage{hyperref}

\begin{document}

\preprint{AIP/123-QED}

\title{Mg$_2$Si is the new black: introducing a black silicide with $>$95\% average absorption at 200-1800 nm wavelengths}

\author{Alexander~Shevlyagin}
\email{shevlyagin@iacp.dvo.ru}
\affiliation{Institute of Automation and Control Processes, Far Eastern Branch, Russian Academy of Science, Vladivostok 690041, Russia}

\author{Vladimir~Il'yaschenko}
\affiliation{Institute of Automation and Control Processes, Far Eastern Branch, Russian Academy of Science, Vladivostok 690041, Russia}

\author{Aleksandr~Kuchmizhak}
\affiliation{Institute of Automation and Control Processes, Far Eastern Branch, Russian Academy of Science, Vladivostok 690041, Russia}
\affiliation{Pacific Quantum Center, Far Eastern Federal University, 8 Sukhanova Str., Vladivostok, Russia}

\author{Eugeny~Mitsai}
\affiliation{Institute of Automation and Control Processes, Far Eastern Branch, Russian Academy of Science, Vladivostok 690041, Russia}

\author{Alexander~Sergeev}
\affiliation{Institute of Automation and Control Processes, Far Eastern Branch, Russian Academy of Science, Vladivostok 690041, Russia}

\author{Andrey~Gerasimenko}
\affiliation{Institute of Chemistry, FEB RAS, 690022, Vladivostok, Russia}

\author{Anton~Gutakovskii}
\affiliation{Institute of Semiconductor Physics of SB RAS, 630090, Novosibirsk, Russia}

\begin{abstract}
Textured silicon surface structures, in particular black silicon (b-Si), open up possibilities for Si-based solar cells and photodetectors to be extremely thin and highly sensitive owing to perfect light-trapping and anti-reflection properties. However, near-infrared (NIR) performance of bare b-Si is limited by Si band gap of 1.12 eV or 1100 nm. This work reports a simple method to increase NIR absorption of b-Si by $in$ $vacuo$ silicidation with magnesium. Obtained Mg$_2$Si/b-Si heterostructure has a complex geometry where b-Si nanocones are covered by Mg$_2$Si shells and crowned with flake-like Mg$_2$Si hexagons. Mg$_2$Si formation atop b-Si resulted in 5-fold lower reflectivity and optical absorption to be no lower than 88\% over 200-1800 nm spectral range. More importantly, Mg$_2$Si/b-Si heterostructure is more adjusted to match AM-1.5 solar spectrum with theoretically higher photogenerated current density. The maximal advantage is demonstrated in the NIR region compared to bare b-Si in full accordance with one`s expectations about NIR sensitive narrow band gap ($\sim$0.75 eV) semiconductor with high absorption coefficient, which is Mg$_2$Si. Results of optical simulation confirmed the superiority of Mg$_2$Si/b-Si NIR performance. Therefore, this new wide-band optical absorber called black silicide proved rather competitive alongside state-of-the-art approaches to extend b-Si spectral blackness.
\end{abstract}

\keywords{black silicon, magnesium silicide, vacuum evaporation, solid phase epitaxy, heterostructures, antireflection, optical absorption, optical modeling, solar energy material.}

\maketitle

\section{Introduction}
Rigid semiconducting properties of the monocrystalline silicon (c-Si) place its single-junction solar cell (SC) efficiency to 29.4\% \cite{veith2018reassessment} in accordance with Schockley-Queisser limit\cite{shockley1961detailed}. Interdigitated back contact approach set a practical milestone of 26.7\% in 2017\cite{yoshikawa2017silicon}. Recently, bifacial Si SC pushed it to approximately 29\% closing the gap between theory and experiment\cite{Url}. Simultaneously, a gradual thinning of Si wafers for reduction of the panel cost takes place. This search for a thinner light absorbing layer is additionally motivated by the much wider flexible SC applicability \cite{hashemi2020recent,um2021flexible,zhao2021recent}. Therefore, modern Si SC industry has been narrowed down to the motto ``thinner \& cheaper`` resulted in 10-fold decrease in thickness (20 $\mu$m) \cite{massiot2020progress} representing ultrathin SC niche and extremely thin ($<$3 $\mu$m) freestanding SC beating 12\% efficiency \cite{xue2020free}. However, the thinner an absorbing layer, the lower short-circuit current density is generated, while generally decrease in Si wafer thickness results in its open-circuit voltage increase. A compromised thickness of 75 $\mu$m is believed to deliver a maximum photoelectric conversion efficiency of the Si SC\cite{chime2022thin}. In addition, absorbing performance can be enhanced by light-trapping effect allowing to maintain thin and ultrathin designs providing higher photogenerated current density in comparison with the flat absorbers\cite{saive2021light}.

Black silicon (b-Si) is a material of choice for the both thickness reduction below 100 $\mu$m toward flexible SC and light absorbing enhancement due to excellent trapping and antireflection properties\cite{otto2015black}. On the other hand, b-Si photovoltaic performance below Si band gap (1.12 eV or 1100 nm) leaves a room for the further improvements. The lower an incident photon energy, the lower optical losses and associated absorbance despite of multiple reflections in black structures. Thus, it is expected that near-infrared (NIR) absorbing layer deposited above b-Si surface would result in extension of its spectral "blackness". So far, b-Si structures have been modified $via$ thin film or nanoparticles deposition, formation of the hierarchical carbon-based nanostructures atop or pulsed laser postprocessing \cite{lu2021enhanced,isakov2018wide,song2020plasmon,wang2021enhancing,sarkar2019photoresponse,shah2016pyrolytic,phan2020cvd,sanchez2021laser,paulus2021obtaining}. To date, practical use of the outlined approaches is limited by the low cost-efficiency factor resulted from the expensive deposition methods applied (atomic-layer deposition, molecular-beam epitaxy, $etc.$) and rare or high melting materials and compounds used (PtS$_2$, NbN, Au, TiN and MoS$_2$).

In this work, we propose a simple way to drastically increase b-Si NIR performance within thin coating of semiconducting (band gap $\sim$0.75 eV \cite{mahan1996semiconducting}) magnesium silicide (Mg$_2$Si) by vacuum evaporation technique. Mg$_2$Si is a Si-compatible optical material with superior absorbance from the ultraviolet (UV) to NIR regions compared to bare Si \cite{kato2011optoelectronic}. It has already demonstrated photovoltaic perspectives as homo-/heterojunction photodetectors and SC \cite{el2019ecofriendly,udono2015crystal,el2019silicon,zhu2021high,yu2021enhanced,shevlyagin2020probing}. Moreover, higher absorption coefficient exceeding 10$^5$ cm$^{-1}$ above 1.5 eV \cite{kato2011optoelectronic} is favorable for Si wafer thinning confirmed by recent Mg$_2$Si/Si SC modeling \cite{deng2017simulation}. In addition, proposed route to cover b-Si with NIR absorbing material is simple, low temperature, scalable and fast method resulted from the benefits of vacuum evaporation.

We show that $(i)$ vacuum evaporation results in b-Si geometry preservation and Mg$_2$Si flakes-like epitaxial growth atop b-Si nanocones, $(ii)$ higher NIR performance is a result of both hierarchical Mg$_2$Si/b-Si structure and Mg$_2$Si intrinsic absorption, $(iii)$ Mg$_2$Si coating resulted in only slight increase in UV-VIS reflection, while pronounced antireflection extends deeply below Si band gap down to that of Mg$_2$Si, and (iv) overall optical performance can be maximized through appropriate silicidation conditions. Optical modeling of the Mg$_2$Si/b-Si structure confirmed better localization of the NIR photons in comparison with uncovered b-Si. Finally, a survey on recent progress in b-Si modification in terms on NIR performance enhancement highlights new competitive black material with averaged over 200-1800 nm absorbance and reflectivity of 96\% and 3.7\%, respectively, called as black silicide.

\section{Experimental details.}

Black Si samples used as initial substrates for silicidation were produced $via$ reactive-ion beam etching (RIE) as described elsewhere\cite{ivanova2013bactericidal}. RIE with SF$_6$ and O$_2$ mixture was performed to produce the nanocones with average height and period of 200 and 100 nm, respectively, on Si(001) wafer.

Next, b-Si samples were cleaned by a standard wafer cleaning process with piranha solution for 10 minutes prior to vacuum chamber loading. For the silicidation of the b-Si a high vacuum turbomolecular pumped evaporation chamber with the base pressure of 1$\times$10$^{-6}$ Torr was used. The chamber is equipped with K-cell for Mg evaporation (5N, Alfa Aesar, USA), quartz crystal microbalance (QCM) sensor and rotating sample's holder with resistive heater. QCM sensor was used for monitoring the deposition rate of Mg source. A calibration factor was obtained by comparing the thickness inferred from the QCM sensor with that of measured by atomic force microscopy on a specially step-shaped film deposited on bare Si(001) substrate. Before silicidation b-Si samples were degassed at 400$^o$C for 20 minutes under the pressure no worse than 8$\times$10$^{-5}$ Torr. A two-step growth technique known as solid phase epitaxy (SPE) was applied for Mg$_2$Si formation. First, 10-300 nm of Mg was deposited at room temperature with evaporation rate of 20 nm/min followed by sample annealing. The former resulted in metallic film formation on b-Si surface with pronounced light reflection contrasting to bare b-Si. It took about 5 minutes at 330-370 $^o$C to obtain film color typical of Mg$_2$Si. During the SPE, vacuum pressure was no worse than 3$\times$10$^{-6}$ Torr. After the sample unloading from the chamber, its morphology, crystal structure, chemical composition and optical properties were investigated.

The X-ray diffraction (XRD) studies were carried out in the 2$\theta$/$\omega$ mode and parallel beam optics geometry (RIGAKU SmartLab diffractometer, Japan). The XRD peaks were identified using the database ICDD PDF-2. Morphology of the Mg$_2$Si/b-Si (black silicide) surface was characterized using scanning electron microscopy (SEM; Ultra 55+, Carl Zeiss, Germany). The crystal structure of prepared Mg$_2$Si/b-Si structure were studied by means of high resolution transmission electron microscopy (HRTEM, TITAN 80-300, FEI Company, USA) operated at 300 kV voltage with point-to-point resolution of 0.165 nm. The energy-dispersive X-ray spectroscopy (EDS) analysis was carried out in the cross-sectional STEM mode, with the electron beam size of $\sim$1 nm. Raman scattering was pumped by a 473 nm CW laser using optical microscopy setup (NTEGRA Spectra II, NT-MDT, Russia) confocally aligned to a grating-type spectrometer with a thermo-electrically-cooled CCD-camera (i-Dus, Oxford Instruments, UK). The hemispheric total reflectance and transmittance for normal incidence in wavelength from 200 to 1800 nm were measured using a spectrophotometer with an integrating sphere (Cary 5000, Varian, USA). For comparison, reference spectra of bare b-Si were also obtained.

Finite-difference time-domain (FDTD) calculations have been carried out using commercial electromagnetic (EM) solver (Lumerical Solutions, Ansys Inc.). The morphology of the Mg$_2$Si/b-Si heterostructure was reproduced from the corresponding cross-sectional STEM image. Local EM field distributions were calculated for the black silicide and pure b-Si structures pumped under normal incidence at 1300 nm wavelengths. Complex dielectric function of the Mg$_2$Si was taken from Ref.\cite{udono2015crystal}.

\section{Results and discussion.}

\begin{figure*}[t!]
\center{\includegraphics[width=0.85\linewidth]{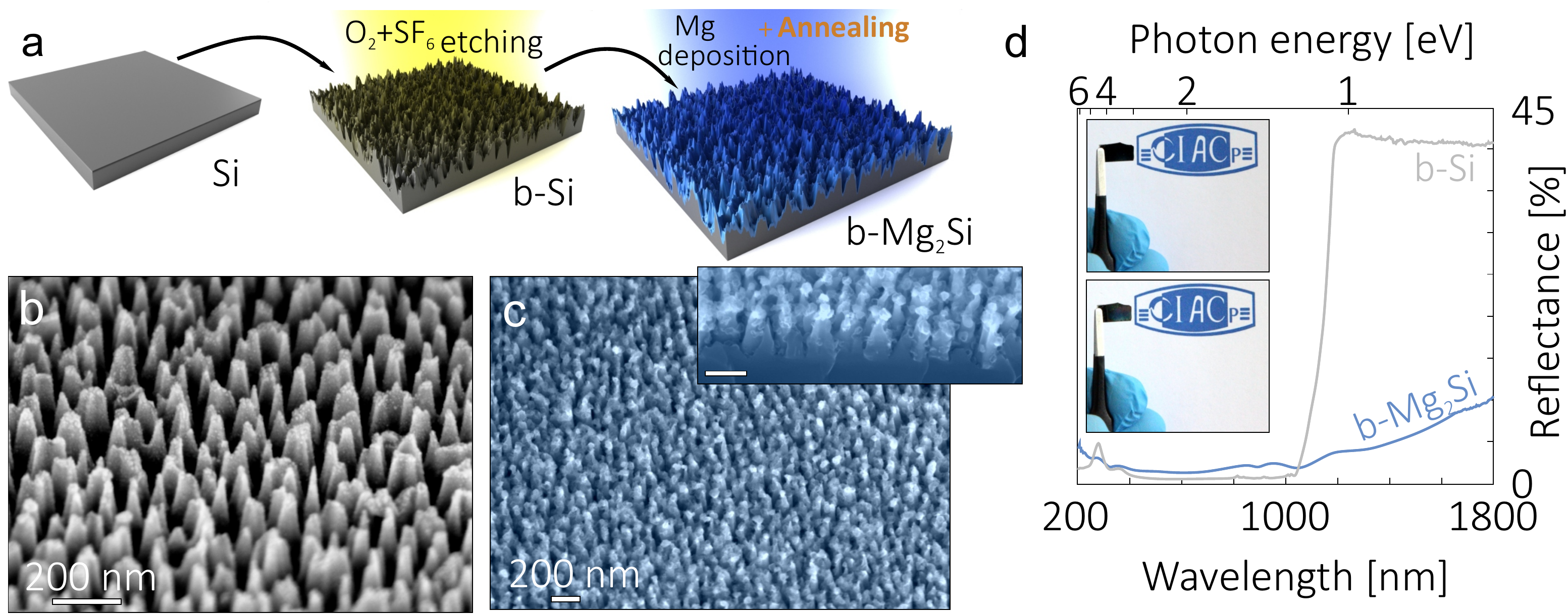}}
\caption{Fabrication of the black magnesium silicide and its optical properties. (a) Schematic illustration of the  b-Mg$_2$Si preparation procedure. Side-view SEM images of the bare (b) and Mg$_2$Si (50 nm) covered (c) surface of the b-Si. Inset shows b-Si nanocones crowned with flake-like Mg$_2$Si faceted hexagons. (d) Reflectance spectra of b-Mg$_2$Si compared to bare b-Si. Insets show photographs of the as-prepared black silicide sample under different tilting angles.
}
\label{fig1}
\end{figure*}

A template substrate for black magnesium silicide preparation was fabricated via reactive ion etching of monocrystalline Si wafer with (001) surface orientation. This procedure resulted in well-known b-Si structure composed of the randomly distributed Si nanocones. Next, vacuum evaporation to cover b-Si surface with Mg$_2$Si layer was applied. Described protocol of the black silicide formation is illustrated in Fig.\ref{fig1}a.

The as-loaded from vacuum chamber sample retains the typical nanocones structure of the bare b-Si shown in cross-sectional SEM images (Fig. \ref{fig1}b and c), whereas deposited Mg$_2$Si layer of the optimal thickness (see discussions below) is presented as flake-like faceted hexagons crowning b-Si edges (inset of Fig. \ref{fig1}c).  The reflectance spectrum of the Mg$_2$Si covered b-Si nanocones array presented in Fig. 1d demonstrates strong NIR antireflection performance when compared to starting b-Si surface. These measurements confirm averaged reflection in the 200-1800 nm spectral range of 3.7\% from black magnesium silicide surface, which is $\sim$5 times lower with respect to 17.6\% of the bare b-Si. However, the former material clearly demonstrates slightly increased VIS reflection with specific silicide related spectral feature at 500 nm (2.5 eV) resulted from first direct interband optical transition in Mg$_2$Si\cite{au1969electronic}. This fact suggests that Mg$_2$Si nanoflakes provide enough area for the surface reflection being the main reason behind slight deterioration of the VIS antireflection. Insets on Figure \ref{fig1}d demonstrate photographs of b-Si specimen after loading from vacuum evaporation chamber where silicidation took place. Grown Mg$_2$Si/b-Si sample looks like uniform black surface to the naked eye. However, clear interface between bare (masked and uncovered with Mg$_2$Si layer) b-Si area and Mg$_2$Si film appears while sample tilting. Moreover, Mg$_2$Si covering layer changes color from the black to deep blue intrinsic for silicide together with slightly higher VIS reflection further confirmed by optical measurements.

\begin{figure*}[t!]
\center{\includegraphics[width=0.7\linewidth]{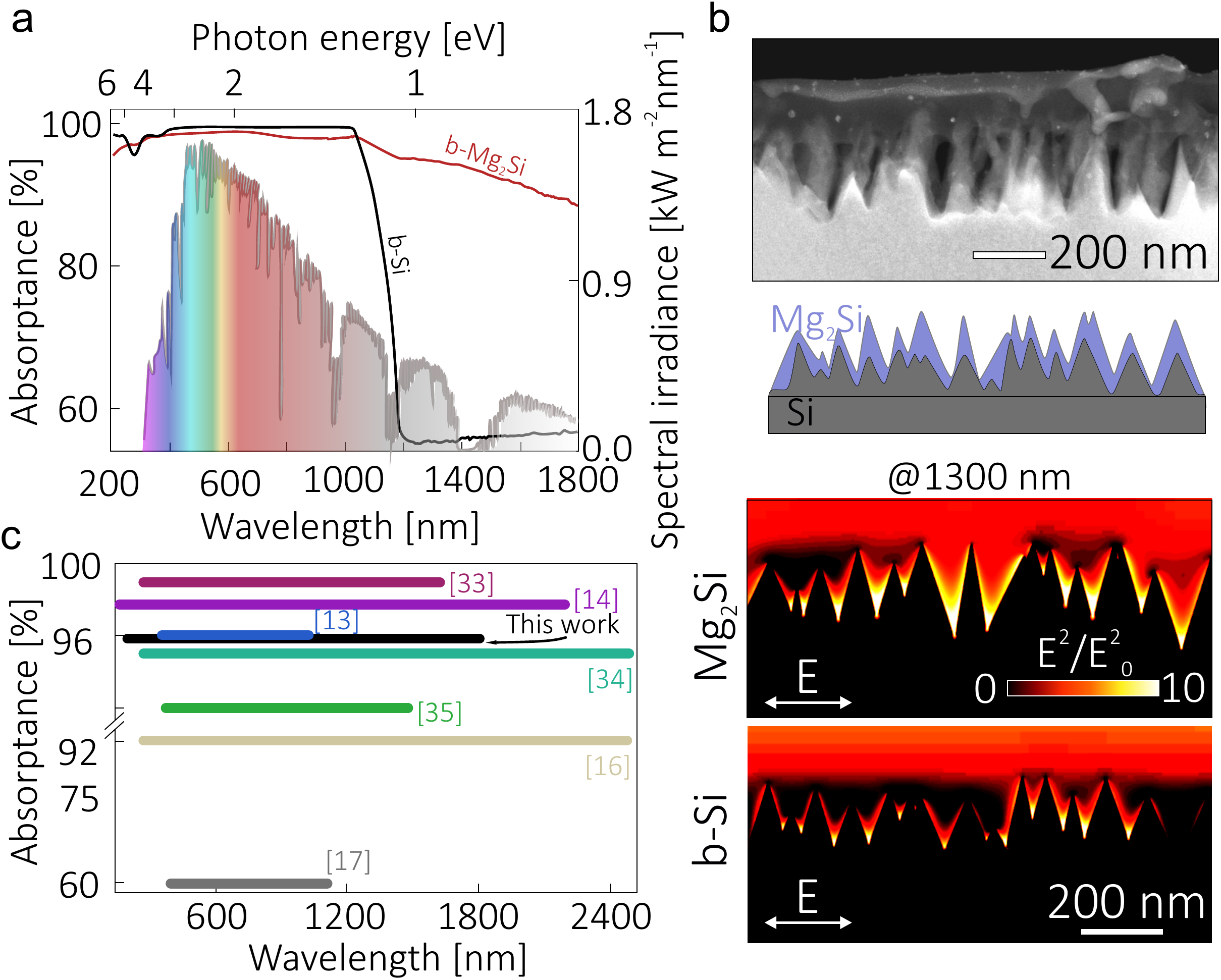}}
\caption{Light absorbing properties of the black silicide in comparison with bare b-Si and other modified b-Si structures. (a) Absorbance spectra of the b-Mg$_2$Si and b-Si calculated from measured reflectance and transmittance with shown solar irradiance standard AM1.5G spectrum. (b) Black silicide morphology reconstruction basing on the STEM HAADF image for modeling normalized squared electric-field amplitude distribution (E$^2$/E$^2_0$) at 1300 nm pump wavelength of the Mg$_2$Si-covered and bare b-Si structures. (c) Survey of operating wavelength range and averaged optical absorption of the produced black silicide and several representative b-Si structures reported in \cite{lu2021enhanced,isakov2018wide,wang2021enhancing,sanchez2021laser,pasanen2020nanostructured,zhong2016enhanced,sheehy2007chalcogen}.
}
\label{fig2}
\end{figure*}

To illustrate the benefits of b-Si with Mg$_2$Si layer in absorbing solar irradiance, optical transmittance (T) and reflectance (R) spectra were recorded from the ultraviolet (UV, 200 nm) to NIR (1800 nm) ranges (Figure \ref{fig2}a) and compared to state-of-the-art black surface structures \cite{lu2021enhanced,isakov2018wide,wang2021enhancing,sanchez2021laser,pasanen2020nanostructured,zhong2016enhanced,sheehy2007chalcogen} (Figure 2c). Optical absorption spectra (A) calculated as 1-T-R demonstrate that optical properties of the b-Si changed drastically after Mg$_2$Si deposition. In general, optical performance at ?$>$1100 nm is significantly reduced in b-Si structures, while black silicide structure can effectively absorb NIR light resulted from the smaller Mg$_2$Si band gap. Moreover, total reflection and absorption normalized to solar irradiance in accordance with expression:

\begin{equation}
R^{norm}_{AM1.5G},A^{norm}_{AM1.5G} = \frac {\int_{\lambda=280 nm} ^{\lambda=1800 nm}  R(\lambda), A(\lambda)\times N(\lambda)d\lambda}{\int_{\lambda=280 nm} ^{\lambda=1800 nm} N(\lambda)d\lambda},
\end{equation}

\noindent where N($\lambda$) is solar flux under AM1.5G illumination, allowed us to conclude that black silicide structure under consideration is superior to bare b-Si in terms of NIR, overall and AM1.5G-normalized performance. Calculated values are summarized in Table 1.

\begin{table*}[]
\caption{Optical performance of the black silicide compared to bare black silicon including optical absorbance, reflectance averaged and normalized over different spectral ranges and photogenerated current densities under AM1.5G solar spectrum irradiance.}
\begin{tabular}{c|c|c|c|c|c|c|c|}
\cline{2-8}
                                          & \begin{tabular}[c]{@{}c@{}}R {[}200-1800   nm{]}, \\ \%\end{tabular} & \begin{tabular}[c]{@{}c@{}}R {[}1100-1800   nm{]}, \\ \%\end{tabular} & \begin{tabular}[c]{@{}c@{}}R AM1.5G,\\  \%\end{tabular} & \begin{tabular}[c]{@{}c@{}}A {[}200-1800 nm{]},\\  \%\end{tabular} & \begin{tabular}[c]{@{}c@{}}A {[}1100-1800 nm{]},\\  \%\end{tabular} & \begin{tabular}[c]{@{}c@{}}A AM1.5G , \\ \%\end{tabular} & \begin{tabular}[c]{@{}c@{}}MAPD, \\ mA/cm$^2$\end{tabular} \\ \hline
\multicolumn{1}{|c|}{b-Si}                & 17.56                                                                & 38.56                                                                 & 29.8                                                    & 81.1                                                               & 58.3                                                                & 92.1                                                     & 50.9                                                    \\ \hline
\multicolumn{1}{|c|}{\textbf{b-Mg$_2$Si}} & \textbf{3.66}                                                        & \textbf{5.87}                                                         & \textbf{2.3}                                            & \textbf{95.8}                                                      & \textbf{92.8}                                                       & \textbf{97.5}                                            & \textbf{54}                                             \\ \hline
\end{tabular}
\end{table*}

Let us pay attention only to the most impressive results obtained. For instance, black silicide structure demonstrates 6-fold decrease in NIR reflection being reduced from 38.5\% down to 5.9\% and 1.5 times higher NIR absorption. Moreover, Mg$_2$Si/b-Si structure is better adjusted to match solar irradiance standard AM1.5G that reflected in 13-fold lower total reflection and higher averaged absorption of 97.5\%. The latter allows to estimate the maximum achievable photo current density (MAPD) of the black silicide structure as:

\begin{equation}
J = \frac{eh}{C}\int_{\lambda=280 nm} ^{\lambda=1800 nm} \lambda \times A(\lambda)\times N(\lambda)d\lambda,
\end{equation}

\noindent where $e$, $h$, $C$ and $\lambda$ are electron charge, Planck constant, velocity of light and photon wavelength, respectively. MAPD reaches 54 mA/cm$^2$ under the assumption of recombination losses absent enhancing b-Si MAPD by 3 mA/cm$^2$. This claims black silicide as solar energy material with enhanced performance beyond 1100 nm.

Next, FDTD calculations were performed to study the effect of b-Si covering with Mg$_2$Si layer. Figure 2b represents geometry of the grown Mg$_2$Si/b-Si heterostructure reconstructed from the HAADF SEM data. The cross-sectional EM field distributions (E$^2$/E$^2_0$) near the surface of black silicide and bare b-Si under illumination are presented in the bottom panels of Figure \ref{fig2}b. The wavelength 1300 nm was chosen to illustrate NIR performance enhancement associated with NIR absorbing Mg$_2$Si cover layer, while bare b-Si can demonstrate marked IR absorption only under heavily doping and/or laser postprocessing \cite{paulus2021obtaining,sheehy2007chalcogen,sher2014picosecond}. Thus, performed optical modeling clearly answers the question concerning increased NIR absorption taking place in black silicide structure. Higher EM field localization and "hot spots" are observed in the case of Mg$_2$Si/b-Si structure with maximum confinement occurred near the b-Si nanocones base and Mg$_2$Si crowns. It can be tentatively concluded that complex hierarchical nanocones-nanoflake structure realized in black silicide resulted in the marked increase of the both NIR antireflection and absorption properties. NIR photons after being back-scattered within b-Si nanocones are effectively absorbed by Mg$_2$Si possessing much higher intrinsic absorption coefficient over wide spectral range.

Significantly, the wide range optical absorption performance of the black silicide is compared with state-of-the-art approaches to enhance NIR efficiency of the b-Si as shown in Fig. \ref{fig2}c. One can see that black silicide delivers absorption of $\sim$96\% over 200-1800 nm spectral range, while it does not require any expensive deposition methods and rare or high melting materials and compounds. For instance, PtS$_2$ \cite{lu2021enhanced} and NbN \cite{isakov2018wide} decorated b-Si with superior absorption were produced under higher temperature budgets, while laser postprocessing of the b-Si often faces with Si amorphization \cite{sanchez2021laser,borodaenko2021deep}, metastable phases formation and low minority carrier lifetime due to metal-like behavior after b-Si hyperdopping\cite{sher2014picosecond,winkler2011insulator,sickel2017microscopic}.

\begin{figure*}[t!]
\center{\includegraphics[width=0.6\linewidth]{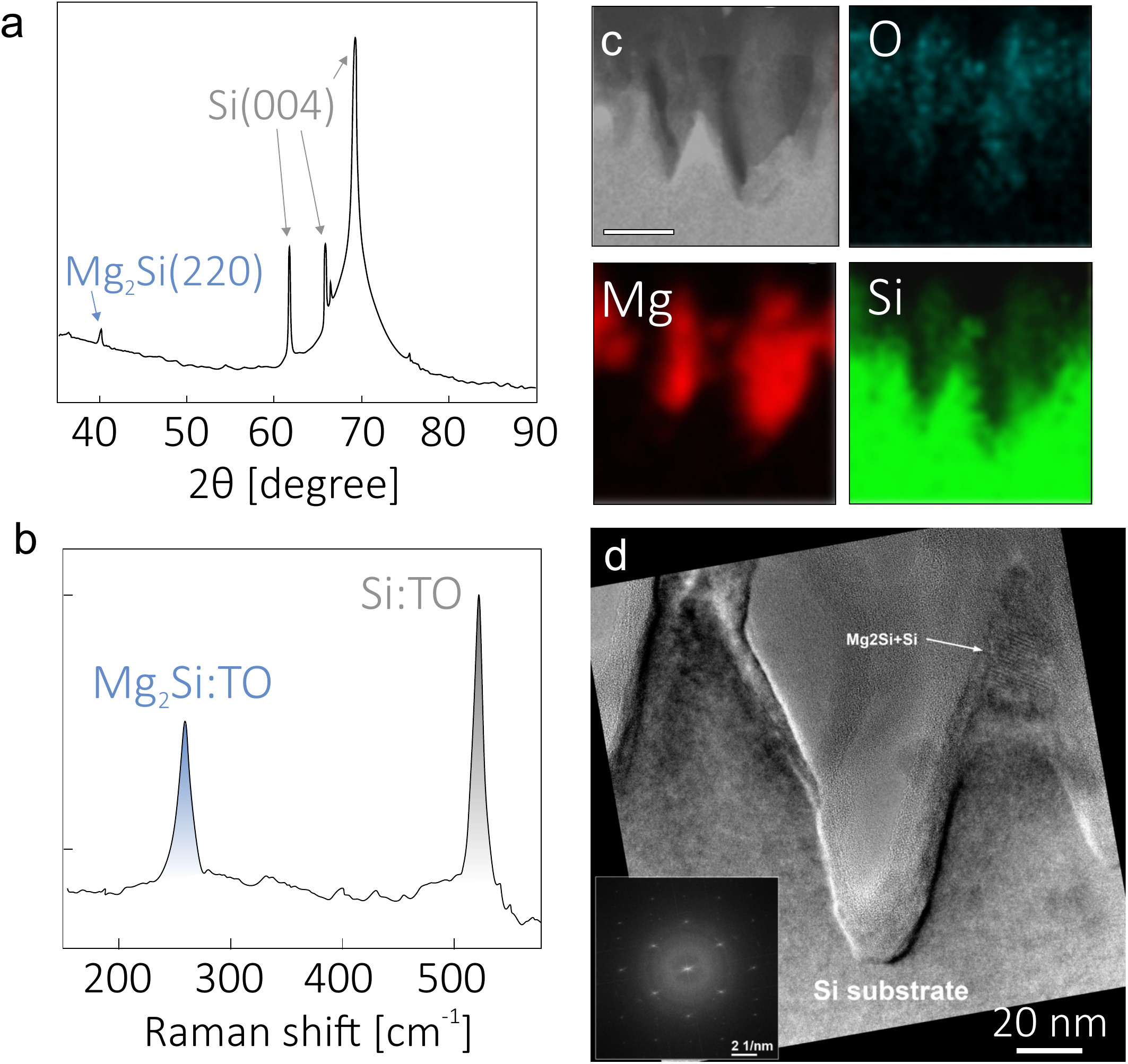}}
\caption{Crystal structure and chemical composition of the prepared black silicide. (a) XRD pattern suggested (220) preferential orientation of the deposited Mg$_2$Si cover layer. (b) Representative Raman spectrum  of the Mg$_2$Si sample showing TO phonon modes of crystalline Si and Mg$_2$Si. (c) Cross-sectional STEM-HAADF image and corresponding chemical elements composition mapping provided by EDS. (d) Representative cross-sectional HR-TEM image demonstrates refined morphology of the Mg$_2$Si deposits, which form core-shell Si-Mg$_2$Si complex nanostructure. Inset shows overall FFT pattern with reflexes corresponding to both Si and Mg$_2$Si.
}
\label{fig3}
\end{figure*}

The second part of our work gives some insights into relation between b-Mg$_2$Si growth conditions and structural parameters as morphology, crystallinity and phase composition. On the other hand, these features strongly affect optical properties of the resulted nanostructures. Therefore, concerned markers were chosen to maximize wide-band antireflection and light absorbing performance of the black magnesium silicide.

Structural and elemental analysis of the black silicide sample are summarized in Figure \ref{fig3}a-d. The XRD pattern and Raman spectra shown in Fig. \ref{fig3}a and b, respectively, suggest that $in$ $vacuo$ b-Si sample cleaning and silicidation at moderate temperature of the solid state reaction ($\sim$330$^o$C) have no influence on b-Si intrinsic properties. XRD and Raman peak positions coincide well with bulk Si and no signs of amorphization are observed. It makes sense in terms of the well-known antireflection properties degradation of the RIE derived b-Si after thermal annealing\cite{crouch2004infrared}. Neither metastable Mg-silicide phases, nor magnesium oxides, hydrates and silicates can be seen regardless low vacuum conditions. In addition, grown Mg$_2$Si layer demonstrates good crystallinity resulting in $(i)$ only (220) diffraction peak appearance corresponded to Mg$_2$Si with Fm3$\overline{m}$ crystal lattice and $(ii)$ pronounced TO phonon mode observation with intensity comparable to that of Si, despite thin silicide film deposition. The former suggests that the grown Mg$_2$Si shells have preferential orientation on the b-Si nanocones. Mg$_2$Si crystallites mean size and residual stress were calculated to be 28 nm and 0.2\%, respectively in accordance with XRD data. Thus, Mg$_2$Si cover layer does not disgrace the best ultrahigh vacuum grown Mg$_2$Si films in terms of crystallinity, whereas much lower vacuum and substrate surface treatment requirements were used \cite{mahan1996semiconducting,shevlyagin2020probing,vantomme1997thin,gouralnik2018formation,gouralnik2021synthesis}. These facts confirm high applicability and scalability degrees of the proposed new black material preparation. HAADF-STEM imaging and EDS elemental mapping (Fig. \ref{fig3}c) have been carried out to illustrate Mg$_2$Si/b-Si morphology and chemical composition. These findings confirm that Mg$_2$Si/b-Si sample is a stacked textured structure with Mg$_2$Si layer repeating b-Si nanocones geometry to some extent. Elemental oxygen is related to atmosphere long-term exposure. A closer look at the individual b-Si nanocones covered with Mg$_2$Si is shown in the HRTEM image (Figure \ref{fig3}d). Magnesium silicide is presented with few-nm thick crystalline shell surrounding b-Si nanocones with well-defined reflexes on FFT inset. Thus, Mg$_2$Si/b-Si material is a hierarchically organized nanostructure where b-Si nanocones are covered with thin Mg$_2$Si shells and crowned by Mg$_2$Si nanoflakes.

\begin{figure*}[t!]
\center{\includegraphics[width=0.9\linewidth]{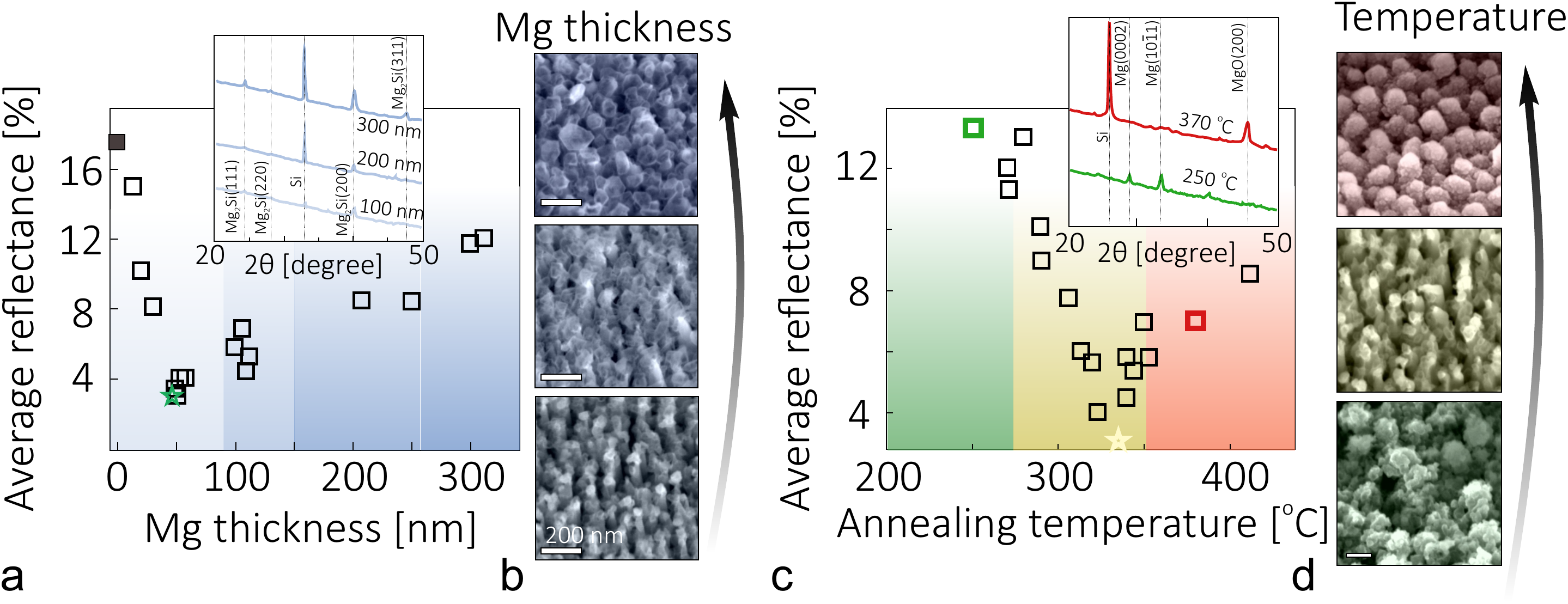}}
\caption{Summary of the optimal Mg film thickness (a, b) and annealing temperature (c, d) findings for boosting NIR antireflection performance of the proposed black magnesium silicide. SEM (b, d) and XRD (insets of a, c) investigations served as morphology and phase composition markers, which have crucial influence on the optical properties. Left panels (a, b) represent samples annealed under the same SPE conditions ($\sim$330$^o$C), while the right ones (c, d) do that of under the constant Mg film thickness of $\sim$50 nm.
}
\label{fig4}
\end{figure*}

To find the optimal conditions for Mg$_2$Si growth atop b-Si, Mg$_2$Si/b-Si samples's crystallinity, morphology, phase composition and optical properties were investigated depending on deposited Mg film thickness and substrate temperature upon solid phase epitaxy. The former varied from 10 to 300 nm and the latter did in 250-450$^o$C temperature range.

When fixing SPE temperature slightly above 300 $^o$C, which is favorable for magnetron sputtering and thermal evaporation of Mg$_2$Si films on Si substrates \cite{el2019silicon,yu2013effects,katagiri2018growth,zhang2017contact}, resulting coated b-Si nanostructures demonstrate non monotonic antireflection behavior (defined as averaged on 200-1800 nm optical reflectance) with increase in Mg layer thickness (see Fig. \ref{fig4}a, main frame). First, one can see an abrupt decrease in reflection as Mg thickness approaches 50 nm being a global minimum with the best antireflection performance. Representative morphology evolution of the Mg$_2$Si/b-Si heterostructures is shown in series of the cross-sectional SEM images in Fig. \ref{fig4}b. Passing global minimum of the Mg$_2$Si/b-Si reflectance, the thicker deposited Mg film, the lower antireflection is demonstrated. This tendency resulted from Mg$_2$Si nanoflakes enlargement and increase in surface reflection from it. At higher thickness almost continuous Mg$_2$Si film is observed. In addition, Mg$_2$Si nanoflakes enlargement resulted in higher intensity of the main Mg$_2$Si (220) and other (200), (111) and (311) diffraction peaks appearance suggested polycrystalline Mg$_2$Si growth atop b-Si nanocones (inset of Fig. \ref{fig4}a). As a result, thin Mg films of about 50 nm in thickness are of interest in terms of the best antireflection properties. It corresponds to formation of a complex Mg$_2$Si/b-Si nanostructures upon SPE annealing, which preserves b-Si light-trapping geometry extending its spectral "blackness".

At the second stage, effect of the SPE temperature on the average reflectivity of the Mg$_2$Si/b-Si was examined for the fixed 50-nm thick Mg film thickness determined previously as optimal. The data are summarized in Fig. \ref{fig4}c showing that minimal reflectance is achieved at $\sim$330 $^o$C. Interestingly, temperature below 270 $^o$C is not enough for complete silicidation of the b-Si nanocones. XRD data (inset of Fig. \ref{fig4}c) clearly demonstrate contribution from metallic hexagonal Mg with (0002) and (10$\overline{1}1$) reflexes. In contrast, when annealing temperature exceeds 370 $^o$C, low vacuum condition used resulted in Mg oxidation rather than silicide formation. It reflects in diffraction from (200) crystal plane of the cubic MgO plane and absence of any Mg$_2$Si related phonon modes in Raman measurements. In addition, in both cases the surface morphology changed to popcorn like structure (Figure \ref{fig4}d) resulting in increase of the total reflectance. Therefore, low temperature ($\sim$330 $^o$C) solid phase epitaxy of the thin ($\sim$50 nm) Mg film atop b-Si structure are both required to modify its optical properties toward NIR "blackness".

\section{Conclusion}
In summary, a simple and low-cost method for boosting NIR absorption performance of black silicon by integrating with magnesium silicide is demonstrated. Resulting Mg$_2$Si/b-Si heterostructure preserves antireflection and light-trapping properties of the b-Si associated with nanocones-like geometry and expands its spectral blackness from 1100 nm to at least 1800 nm owning to NIR sensitive narrow-band-gap Mg$_2$Si cover layer. Explored and verified by modeling optical properties of this new wide-band black material called black silicide suggest some possible applications ranging from NIR optical absorbers, high-sensitive NIR devices and solar cells. Results obtained propose material scientists, especially dealing with Si-silicide tandem solar cells engineering, a simple technology to enhance photoelectric conversion efficiency using a b-Si as a platform for silicidation with other environment-friendly and Si-compatible silicides, for example BaSi$_2$ or $\beta$-FeSi$_2$.


\section{Declaration of Competing Interest }
The authors declare that they have no known competing financial interests or personal relationships that could have appeared to influence the work reported in this paper.

\section{Acknowledgements}
This research was supported by the Russian Science Foundation under Grant No. 20-72-00006. Authors are thankful to Prof. Saulius Juodkazis from Optical Sciences Centre and ARC Training Centre in Surface Engineering for Advanced Materials, School of Science, Swinburne University of Technology for providing bare black Si samples.

\section{References}

\end{document}